# A
# 3D RGB Axis-based Color-oriented Cryptography

**Document version:** 1.0
**Document author:** Kirti Chawla
**Document signature:** 07072005_1023:20914

# Introduction

In this document, a formal approach to encrypt, decrypt, transmit and receive information using colors is explored. A piece of information consists of set of symbols with a definite property imposed on the generating set. The symbols are usually encoded using ASCII scheme. A linear-to-3D transformation is presented. The change of axis from traditional [x, y, z] to [r, g, b] is highlighted and its effect are studied. A point in this new axis is then represented as a unique color and a vector or matrix is associated with it, making it amenable to standard vector or matrix operations. A formal notion on hybrid cryptography is introduced as the algorithm lies on the boundary of symmetric and asymmetric cryptography. No discussion is complete, without mentioning reference to communication aspects of "secure" information in a channel. Transmission scheme pertaining to light as carrier is introduced and studied. Key-exchanges do not come under the scope of current frame of document.

# Information and Encoding Scheme

In this section, a basic notion of information and encoding scheme is introduced. Perceive-able information is created from a set of symbols having a pre-defined property imposed on all set-members. The scope of information is limited to set of symbols, which consists of numbers, english characters (small, capital) and a few special symbols. These symbols together form a set called ASCII set. There are 256 symbols in this set. Total bits required to represent the whole set is $LOG_2[256]$ bits or 8-bits. Let's introduce a formal notion to represent this set as a list, where members can be accessed using indices. This can be seen as under:

$$S = \{x \mid ((x \in A_1) \lor (x \in A_2) \lor (x \in A_3) \lor (x \in A_4)) \land (\Delta(x) \leq L)\}$$

Where, the symbols have the following meaning:

S = A set which has ASCII characters as members
x = A set member
$A_1$ = A sub-set of whole numbers limited in range [0, 9]
$A_2$ = A set of small english characters
$A_3$ = A set of capital english characters
$A_4$ = A set of special characters
$\Delta$ = Number-of-bits operator
L = $LOG_2[256]$ or 8

The member of this set can be accessed using indices as in an ordered list. This notion is introduced as under:

$$\begin{bmatrix} x(1) \\ x(2) \\ \vdots \\ x(n) \end{bmatrix} = \begin{bmatrix} Member_1 \\ Member_2 \\ \vdots \\ Member_{2^L} \end{bmatrix}$$

where, [x(1) ... x(n)] represent symbols with a range from [1, 256]. Each member has a unique L-bit code in this list. It can be seen as a "linear" sequence of bits or as a (L-1)-degree polynomial. Alternatively a member with L-bit code as [10110100] can be seen, $f(x) = x^7 + x^5 + x^4 + x^2$. Operations like addition and multiplication are always *modulo-$2^L$* so as to wrap around in case of overflows, hence multiplication of [x(170)] = [10101010] and [x(241)] = [11110001] is given as [x(2)] = [00000010]. This also gives the notion of closure property over set members. Alternative base-system like hexadecimal or octal can be used, for the sake of brevity, hence may mean $[x(170)_{16}]$ = [AA] and $[x(170)_8]$ = [252] respectively. The property can be extended over other operators as well. The notion of *modulo-$2^L$* can also be extended over other bases.

# Linear-to-3D Transformation

In this section, the notion of Linear-to-3D transformation is introduced. The notion of linear sequence of bits was introduced in previous section. The transformation which maps a linear sequence of bits to 3D is given as under:

$$T[x\{i\}] = \begin{cases} G_{012} & ; \text{X-axis} \\ G_{345} & ; \text{Y-axis} \\ G_{678} & ; \text{Z-axis} \end{cases}$$

Where, the symbols have the following meaning:

T = Linear-to-3D transformation
$x(i)$ = A set member of set S with index $\{i\}$
$G_{012} = B_0B_1B_2$ {Bits 0-2 of $x(i)$, where bit-0 is added for symmetry purposes}
$G_{345} = B_3B_4B_5$ {Bits 3-5 of $x(i)$}
$G_{678} = B_6B_7B_8$ {Bits 6-8 of $x(i)$}

The linear sequence is divided into groups of 3-bits (one extra bit is pre-pended for symmetry purposes and is always 0) and mapped onto 3D as a point P[x, y, z], where each of [x], [y] and [z] consists of $G_{012}$, $G_{345}$ and $G_{678}$ respectively. Each of [x], [y], [z] has N-bits inclusive of their respective group bits, where N > 3. More number of bits in each of [x], [y] and [z], results into more points in the space.

Extending the notion of *modulo-$2^L$* to N-bit per dimension space results into *modulo-$2^N$* based operations. Each of the *modulo-$2^N$* operation (addition, subtraction, multiplication …) can be applied individually per dimension.

The whole concept can be visualized as the following illustration.

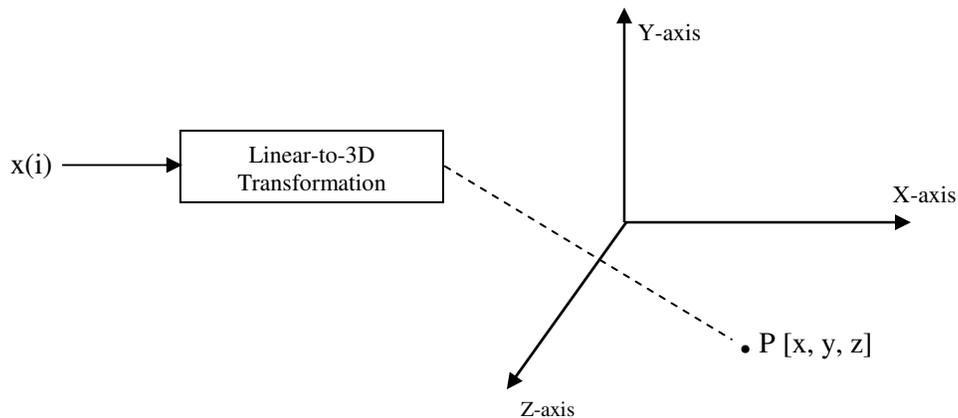

# Axis Transformation

In this section, the traditional 3D axis is transformed into a different axis but with same number of dimensions. Individually [x], [y] and [z] are mapped onto [r], [g] and [b] axis respectively. So a point P[x, y, z] in traditional space is transformed to P[r, g, b]. This also introduces a subtle property called color. A unique point P[r, g, b] represents a unique color. The number of bits per dimension in RGB axis results into more number of points and hence more number of colors. Extending *modulo-$2^N$* operations here results into a definite color per transformation. The following transformation shows the illustration of the whole concept:

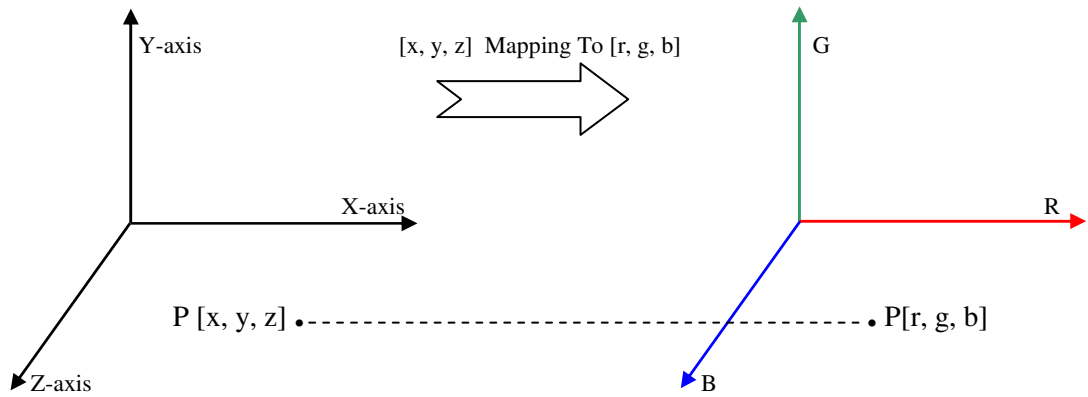

The transformation described above is an implied operation. The notion of change of axis even though denotes a point P in RGB axis; the output is set of bits. These set of bits are used to influence the color encoding in a beam of light, which acts as carrier. Total number of points or colors in a RGB axis with N-bits per dimension is given as under:

$$C = 2^{3N}$$

Where, the symbols have the following meaning:

C = Total number of colors or points in RGB axis
N = Total number of bits in each dimension {r}, {g} and {b}

A point P in RGB axis represents a vector or matrix of dimensions (1 x 3). This vector has to be homogeneous in nature for the sake of operations to be consistent across all dimensions. This introduces the notion of identity element. The transformed vector or matrix hence can be seen as under:

$$V = P[r,g,b,1]$$

Various operations can now be applied onto this transformed vector or matrix. For the sake of brevity P is dropped from the equation and simply can be stated as V = [r, g, b, 1]. The symbol [x(i)] represents a unique point, a unique vector or matrix and a unique color in RGB axis for a unique index [i].

# Point Operations

In this section, various vector or matrix operations are explored, which can be applied and its effects are studied. Continuing from previously introduced notion of vector or matrix as V in RGB axis, extensions are now presented to operate on this vector or matrix.

Let V be denoted as [V] for sake of complete-ness. The dimension of [V] is (1 x 4), inclusive of identity element, which is required when performing homogeneous operations. Any binary operation (addition, subtraction, multiplication …) can be applied with the use of dimensionally-matching vector or matrix. A few basic operations are given as under:

1. **Addition:** In this operation, the dimension of second vector or matrix [K] has to be equal to [V]. The symbolic representation of addition operation is given as under:

$$[O]_{1\times 4} = [V]_{1\times 4} + [K]_{1\times 4}$$

2. **Subtraction:** In this operation, the dimension of second vector or matrix [K] has to be equal to [V]. The symbolic representation of subtraction operation is given as under:

$$[O]_{1\times 4} = [V]_{1\times 4} - [K]_{1\times 4}$$

3. **Multiplication:** In this operation, the number of rows in second vector or matrix [K] has to be equal to number of columns of [V]. The symbolic representation of multiplication operation is given as under:

$$[O]_{1\times j} = [V]_{1\times 4} \times [K]_{4\times j}$$

Binary operations, as described as above, can be performed any number of times. A general notation for all such operations is given as under:

$$[O]_{a_1 \times a_N} = (([V]_{a_1 \times a_2} \odot [K_1]_{a_2 \times a_3}) \odot ... [K_N]_{a_{N-1} \times a_N})$$

This essentially means, a given point in RGB axis can be manipulated as a vector or matrix by applying a series of operations. The effect of these operations is potential relocation of the point. The following illustration shows the concept:

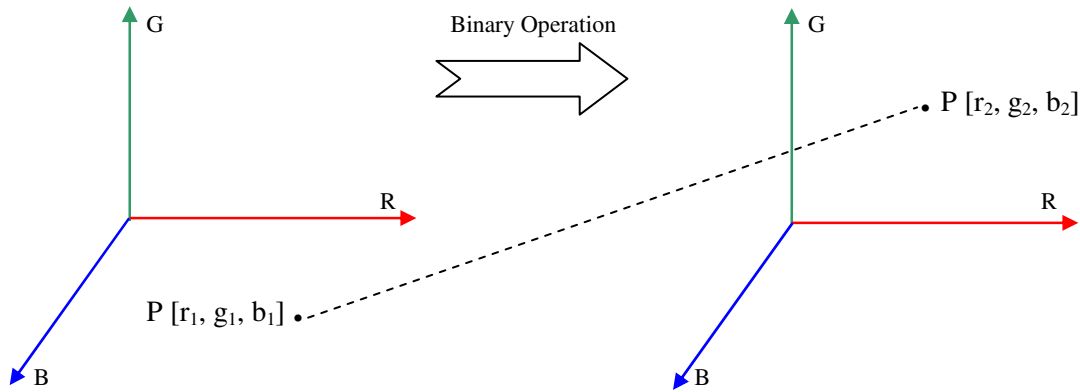

# Cryptography in 3D

In this section, point operations introduced in previous section are explored for cryptographic purposes. A defines crypto-system defines at-least two operations, which are given as under:

1. **Encryption:** In this operation a given symbol is transformed into another symbol by a set of operation(s).

2. **Decryption:** In this operation a given symbol is transformed back to original symbol by a set of operation(s).

The notions of symbols and operations are already defined in previous sections. Extension to previously defined descriptions, from the perspective of cryptography, is given as under:

1. **Message:** A message is juxtaposition of symbols from set S. A symbol is a vector or matrix as defined in previous section.

2. **Key:** A key is secret datum which sender uses to operate upon message using encryption operation. The receiver gets the encrypted message and applies decryption operation to decrypts it using his key to get the original message. A key is a vector or matrix.

3. **Sender:** A sender sends a message to receiver.

4. **Receiver:** A receiver accepts the message from sender.

5. **Channel:** A link over which sender and receiver communicate.

With these preliminaries defined, the encryption and decryption operation can now be defined. They are defined as under:

**Encryption:** Let message be defined as M, where M is juxtaposition of symbols from set S. It can be represented as under:

$$M = x(1) \cdot x(2) \cdot ... x(n)$$

The point operations defined in previous sections can now be applied to above message to encrypt it. Let the operation be defined as O[x, k], where [x] is a symbol from above message and [k] is a key. The effect of this operation is movement of point in RGB axis to a new location. Cumulatively, the encryption operation can be seen as under:

$$E(M) = O[x(1), k(1)] \cdot O[x(2), k(2)] \cdot ... O[x(n), k(n)]$$

Where, the symbols have following meaning:

E = Encryption operation
M = Message
O[x, k] = operation
x(1)…x(n) = Symbols from set S
k(1)…k(n) = Keys

The effect of this operation is that individually [x(1), x(2), …x(n)] are relocated to new position in RGB axis with the help of keys [k(1), k(2), …k(n)]. Alternatively, one can move a set of symbols by fixed or variable amount in RGB axis, hence making it amenable for stream and block of symbols. Operations that are suitable for relocation of points in RGB axis, are vector or matrix (addition, subtraction, multiplication) oriented. It is computationally intensive to find the starting locations of each of these symbols, after they have relocated and keys are kept secret.

**Decryption:** Let encrypted message be defined as M', where M' is juxtaposition of encrypted symbols from set S. It can also be represented as under:

$$M' = y(1) \cdot y(2) \cdot ... y(n)$$

The point operations defined in previous sections can now be applied to above message to decrypt it. Let the operation be defined as O[y, l], where [y] is a encrypted symbol from above message and [l] is an inverse key. The effect of this operation is movement of point in RGB axis to a previously known location. Cumulatively, the decryption operation can be seen as under:

$$D(M') = O[y(1), l(1)] \cdot O[y(2), l(2)] \cdot ... O[y(n), l(n)]$$

Where, the symbols have following meaning:

D = Decryption operation
M' = Encrypted Message
O[y, k] = operation
y(1)…y(n) = Encrypted symbols, where y(i)=O[x(i), k(i)]
l(1)…l(n) = Inverse keys, where l(i) = k(i)$^{-1}$

The effect of this operation is that individually [y(1), y(2), …y(n)] are relocated to original position in RGB axis with the help of inverse keys [l(1), l(2), …l(n)]. Alternatively, one can move a set of symbols by fixed or variable amount in RGB axis, hence making it amenable for stream and block of symbols. Operations that are suitable for relocation of points in RGB axis, are vector or matrix (addition, subtraction and multiplication) oriented.

**Hybrid Cryptography:** The algorithm given here lies in between symmetric and asymmetric cryptography in terms of key used. The keys need to be transported to receiver as in the case of symmetric cryptography, but the keys that are used to decrypt the message are not the same as the ones which are used to encrypt and hence behaving more like asymmetric cryptography. This scheme can be seen with the following illustration:

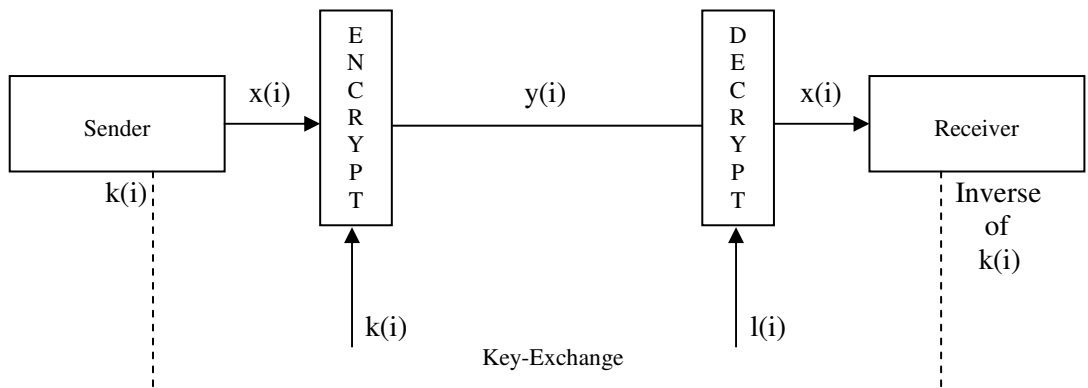

# Transmission

In this section, transmission aspects of information are discussed. The symbol is represented as a vector or matrix on both sender and receiver side. The elements in a given vector or matrix provide not only location information in 3D space, but also color information in RGB axis. Individually, [x], [y], [z] can be input to a light encoder, which encodes the color information in the carrier, to be decoded by the receiver. Each "packet" consists of interleaved color and other meta-data information. There can be two approaches to organize data. First approach requires sender and receiver also share a precise clock or counter and hence "synchronize" per color information. Second approach requires end-of-packet marker to indicate boundaries between color information. Besides this classification of color information, data can also be grouped in set of symbols to increase information content per packet. A generic model of transmission is presented as follows:

**Generic Model of Transmission:** In this model, basic building blocks of transmission, which take advantage of color information are presented. The components required are given as under:

1. *Sender*: A sender generates encrypted symbols.
2. *Receiver*: A receiver consumes decrypted symbols.
3. *Light-Encoder*: A light-encoder takes encrypted symbols and encodes it into light.
4. *Light-Decoder*: A light-decoder takes encoded light signal and generates encrypted symbols.
5. *Transmission Channel*: A medium in which symbols are transmitted.

The illustration of generic model of transmission with all the above components and the components described in previous section are given as under:

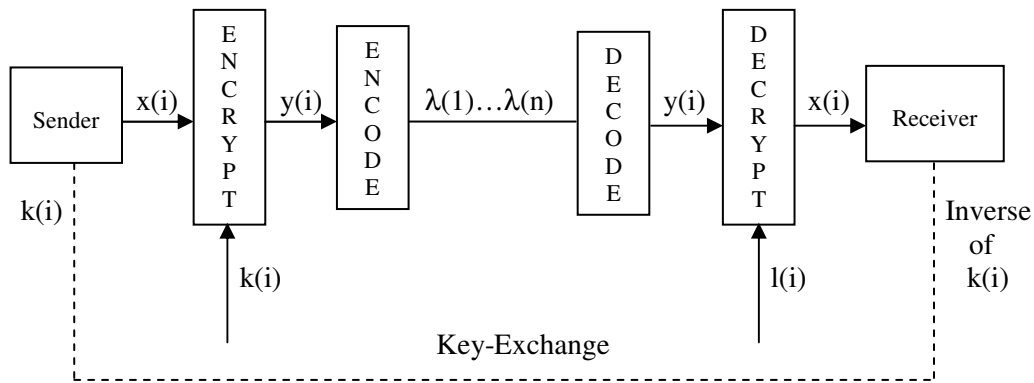

Where, the terms have following meaning:

$x(i)$ = A symbol from set S
$y(i)$ = Encrypted symbol
$\lambda(1)\ldots\lambda(n)$ = Channel symbols
$k(i)$ = Key
$l(i)$ = Inverse key

Encoding or modulation of encrypted symbol results into channel symbol.

**Synchronous Transmission:** In this mode of transmission, sender and receiver share a clock or counter, which keep both of them in-sync with respect to symbols. The illustration showing this concept is shown as under:

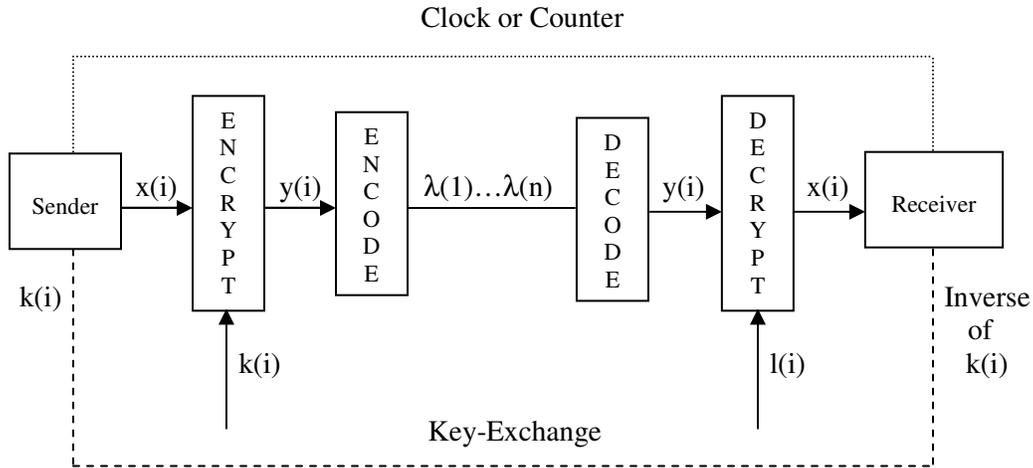

The format of packet depends upon color information and meta-data information. There can be two possible in-order arrangements of the above-said information. In one arrangement the color information is interleaved with meta-data information in fine-grain manner, whereas in other arrangement the interleaving is more coarse-grain. The formats of these arrangements are shown as under:

1. *Fine-grain interleaving format*: In this format, the color information is more closely interleaved with meta-data information. The format is shown as under:

   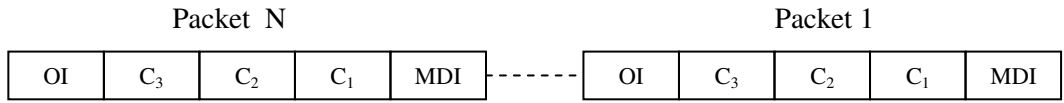

   Where, the symbols have following meaning:

   MDI = Meta-Data Information {Clock or Counter, Packet-Number …}
   $C_1…C_3$ = Color Information {Encrypted Symbol as vector or matrix}
   OI = Other Information {Wrap-around factor, Bits-per-axis …}

2. *Coarse-grain interleaving format*: In this format, the color information is coarsely interleaved with meta-data information. The information is shown as under:

   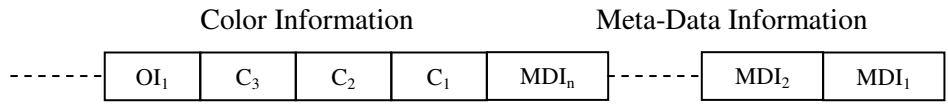

   Where, the symbols have following meaning:

   $MDI_1…MDI_n$ = Meta-Data Information {Clock or Counter, Packet-Number …}
   $C_1…C_3$ = Color Information {Encrypted Symbol as vector or matrix}
   $OI_1$ = Other Packet-specific Information {Wrap-around factor, Bits-per-axis …}

**Asynchronous Transmission:** In this mode of transmission, sender and receiver do not share a clock, but rely on end-of-packet markers, which keep both of them in-sync with respect to symbols. The illustration showing this concept is shown as under:

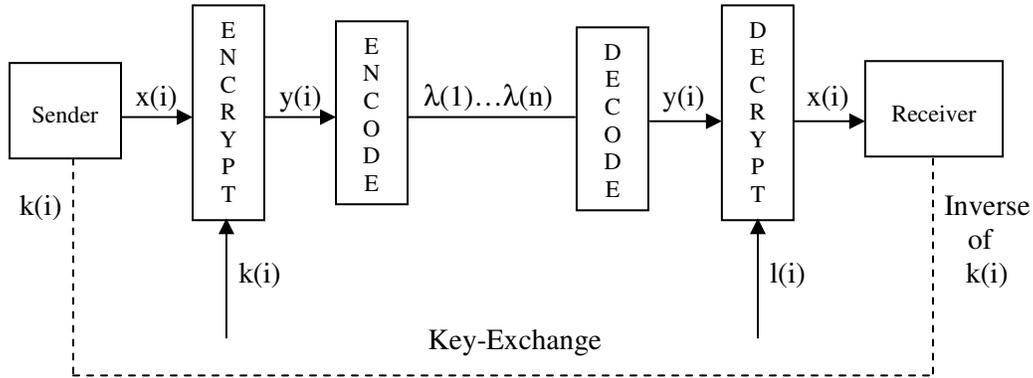

The format of packet depends upon color information and end-of-packet marker. There can be two possible in-order arrangements of the above-said information. In one arrangement the color information is interleaved with end-of-packet marker in fine-grain manner, whereas in other arrangement interleaving is more coarse-grain. The formats of these arrangements are shown as under:

3. *Fine-grain interleaving format*: In this format, the color information is more closely interleaved with end-of-packet marker. The format is shown as under:

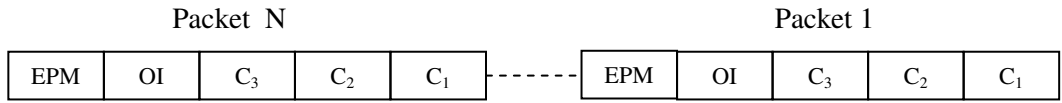

Where, the symbols have following meaning:

$C_1 \ldots C_3$ = Color Information {Encrypted Symbol as vector or matrix}
OI = Other Information {Wrap-around factor, Bits-per-axis …}
EPM = End-of-Packet Marker

4. *Coarse-grain interleaving format*: In this format, the color information is coarsely interleaved with end-of-packet marker. The information is shown as under:

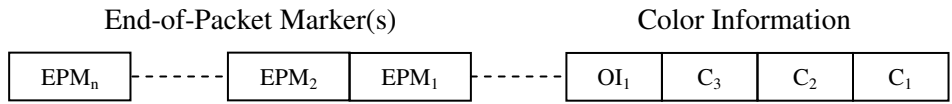

Where, the symbols have following meaning:

$C_1 \ldots C_3$ = Color Information {Encrypted Symbol as vector or matrix}
$OI_1$ = Other Packet-specific Information {Wrap-around factor, Bits-per-axis …}
$EPM_1 \ldots EPM_n$ = End-of-Packet Marker(s)

The useful information per packet can be determined using Information-to-Packet Ratio.

# Appendix

**Wrap-around:** This phenomenon happens when any of the chosen binary operation (Addition, Subtraction, Multiplication …) results into values greater than defined by number of bits. In this case, a method is introduced which ensures re-construction of value. The preliminaries are defined as under:

$$N = (2^L \times WAF) + R$$

Where, the symbols have following meaning:

N = Number to be re-constructed
L = Total bits allocated per number
WAF = Wrap-Around-Factor or dividend of division of N by $2^L$
R = Remainder from division of N by $2^L$

Consider the concept when two numbers are manipulated using binary operation. The details are given as under:

$$T = N_1 \odot N_2$$
$$T > 2^L$$
$$R = T \bmod 2^L$$
$$WAF = \frac{T}{2^L}$$

Where, the symbols have following meaning:

T = Result of binary operation on $N_1$ and $N_2$
$N_1$, $N_2$ = L-bit numbers

The result of this operation is such that it exceeds the maximum possible value given L-bits. Remainder is stored as vector or matrix in 3D space. Wrap-around-factor is stored with the vector information on per symbol or set of symbol basis.

**Bits-per-axis:** This metric is used to provide information about the total number of bits allocated per axis, when re-constructing the encrypted symbol. During the binary transformation, there is a possibility that the total number of bits exceed the initial grouping of 3-bits per axis, hence to accommodate this behavior, special organization of bits can be allowed. Bits-per-axis or BPA is 3-tuple and is given as BPA = ($T_R$, $T_G$, $T_B$). The following illustration shows the concept:

| $T_B$ | $T_G$ | $T_R$ |
|---|---|---|

Where, the symbols have following meaning:

$T_R$, $T_G$, $T_B$ = Total number of bits in encrypted symbol in RGB-axis

This information can be transmitted alongside the encrypted symbol over the channel and poses no threat, infact acts as a "lure" to attacker for re-constructing the encrypted symbol.

**Information-to-Packet Ratio:** A given packet contains color information and other relevant information. This other relevant information can be collectively called a header. This ratio helps to determine color information in a packet vis-à-vis its header. The ratio is given as under:

$$IPR = \left(\frac{B_C}{B_P}\right)$$

Where, the symbols have following meaning:

IPR = Information-to-Packet Ratio
$B_C$ = Bytes allocated for color information
$B_P$ = Bytes allocated for total packet, including header

The ratio gives the amount of information present per packet or actual packet utilization.

**Configurability:** The concept can be configured in multitude of ways. Some of the ways are given as under:

1. Bits-per-axis during linear-to-3D transformation
2. Bits-per-axis during re-construction
3. Total number of binary transformations applied per symbol
4. Fixed or variable binary transformations per axis
5. Packet format and utilization
6. Encoding and decoding schemes for light
7. Other information added as header and trailer in packet

There may be other means to configure various aspects of the concept, which are not mentioned here. The concept is amenable both of stream of symbols or block of symbols. In later case, a set of points in 3D space represent the block of symbols and hence can be manipulated individually or in combined manner.

**Algorithm:** The algorithm for the whole concept can be divided into two sections. First section is dedicated to encryption operation and second section is dedicated to decryption operation. Algorithms pertaining to pertinent operations are given as under:

1. *Encryption operation:* This section defines the algorithm pertaining to encryption operation. It is given as under:

    **Step 1:** Take a symbol [x(i)] or set of symbols
    **Step 2:** Transform the symbols into 3D space {linear-to-3D transformation}
    **Step 3:** Treat 3D axis as RGB axis
    **Step 4:** Let [V(i)] be the homogeneous vector associated per symbol [x(i)]
    **Step 5:** Treat [V(i)] as a point in RGB axis
    **Step 6:** Apply any number of binary transformations to the point, with the help of key vectors [K(i)] {translation, rotation …}
    **Step 7:** Keep track of overflows and maintain wrap-around-factor per symbol
    **Step 8:** Output the symbol [y(i)] or set of symbols
    **Step 9:** Keep track of bits-per-axis for re-construction
    **Step 10:** Stop

2. *Decryption operation:* This section defines the algorithm pertaining to decryption operation. It is given as under:

    **Step 1:** Take a symbol [y(i)] or set of symbols
    **Step 2:** Transform the symbols into 3D space {linear-to-3D transformation}
    **Step 3:** Treat 3D axis as RGB axis
    **Step 4:** Let [V'(i)] be the homogeneous vector associated per symbol [y(i)]
    **Step 5:** Treat [V(i)] as a point in RGB axis
    **Step 6:** Apply same number of binary transformations to the point, with the help of inverse key vectors [L(i)] {translation, rotation …}
    **Step 7:** Take into account wrap-around-factor and bits-per-axis during re-construction of symbol
    **Step 8:** Output the symbol [x(i)] or set of symbols
    **Step 9:** Stop

**Strength:** The strength of algorithm lies in the fact that given a point in 3D space, it is difficult to find its earlier position, if the transformation information is not given. The transformation forms the basis of keys. The algorithm may reveal the nature of transformations, without revealing the values of those transformations, hence making it difficult to find out the exact previous location of a point in 3D space, which actually is a symbol in linear space. This concept can be seen as under:

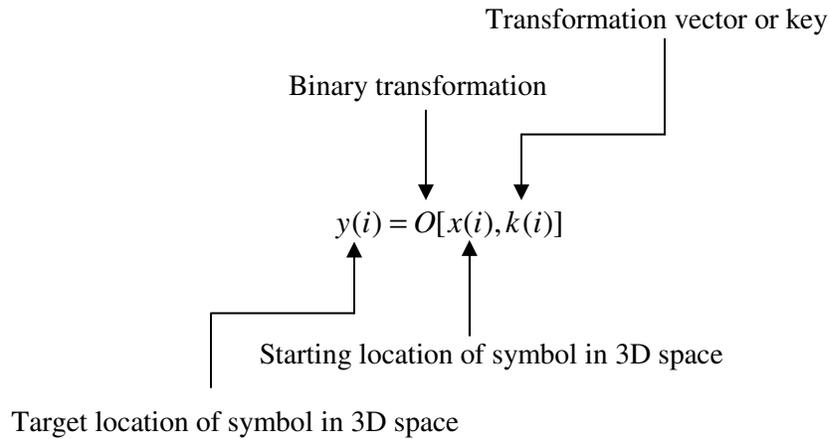

As already been discussed, the symbols are transformed to points in 3D space from linear space, if transformation vector or key in not known and only [y(i)] is given at an instant, the problem is to find [x(i)], with only the given information. Alternatively, the strength also comes from the fact that the symbol present on channel may not directly correspond to encrypted symbol, so the attacker first has to re-construct the encrypted symbol and then transform it to original symbol.

**Example:** The example shown here considers set of three symbols and applies a single binary transformation on all of them. The details are given as under:

Let the set of symbol be denoted as S = {'A','0','/'}, where [x(0)] = 'A', [x(1)] = '0' and [x(2)] = '/'. The equivalent binary values are [x(0)] = [01000001], [x(1)] = [00110000] and [x(2)] = [00101111].

Linear-to-3D transformation for the above symbols is given as [x(0)] = {001, 000, 001}, [x(1)] = {000, 110, 000} and [x(2)] ={000, 101, 111} respectively. Each of these symbols now corresponds to a unique point in 3D space. There are 8-bits per axis, of which 3-bits per axis is symbol information.

Above points in 3D space can be represented in the form of homogeneous vectors and are given as under:

$$[X(0)] = [1\ 0\ 1\ 1],\ [X(1)] = [0\ 6\ 0\ 1],\ [X(2)] = [0\ 5\ 7\ 1]$$

Let the binary transformation be multiplication and the value of binary transformation or vector is given as under:

$$[V] = \begin{pmatrix} 1 & 0 & 0 & 0 \\ 0 & 1 & 0 & 0 \\ 0 & 0 & 1 & 0 \\ 3 & 4 & 5 & 1 \end{pmatrix}$$

The result of the binary transformation is translation of points to a new location. This can be seen as under:

$$[Y(0)] = [4\ 4\ 6\ 1],\ [Y(1)] = [3\ 10\ 5\ 1],\ [Y(2)] = [3\ 9\ 12\ 1]$$

Once the points are relocated, they can be transformed back from 3D space to linear space. This can be seen as [y(0)] = {100, 100, 110}, [y(1)] = {011, 1010, 101} and [y(2)] = {011, 1001, 1100} respectively. When all the bits are combined, it can be represented as [y(0)] = [100100110], [y(1)] = [0111010101] and [y(2)] = [01110011100] respectively. As it can be inferred from the values that they exceed $2^L$, where L = 8, hence modulo-$2^L$ approach is taken to wrap-around. The wrap around values are given as [z(0)] = [00100110], [z(1)] = [11010101] and [z(2)] = [10011100] respectively. The wrap-around-factor for each of the symbol is given as [w(0)] = [00000001], [w(1)] = [00000001] and [w(2)] = [00000011] respectively. Bits-per-axis for each of the symbol is given as [b(0)] = [333], [b(1)] = [343] and [b(2)] = [344] respectively. The encrypted symbols are [y(0)], [y(1)] and [y(2)] respectively and further "mangling" with wrap-around-factors and bits-per-axis results into [z(0)], [z(1)] and [z(2)] respectively, which is transmitted over channel. Wrap-around-factors and bits-per-axis are transmitted alongside the "mangled" symbols over the channel. On the receiver end, the symbols [z(0)], [z(1)] and [z(2)] are "re-constructed" with the help of wrap-around-factors and bits-per-axis and plotted in 3D space, where the inverse binary transformation or inverse key is applied to get the original set of symbols.